\documentclass[11pt]{article}
\usepackage{amssymb,amsmath,amsfonts}
\usepackage{graphicx}
\usepackage{graphics}
\usepackage{eepic,epsfig}

\begin{document}

\title{Vacuum fluctuations and topological Casimir effect in
Friedmann-Robertson-Walker cosmologies with compact dimensions}
\author{ A. A. Saharian$^{1,2}$\thanks{%
E-mail: saharian@ictp.it } \thinspace\ and A. L. Mkhitaryan$^{1}$ \vspace{%
0.5cm} \\
\textit{$^1$Department of Physics, Yerevan State University,} \\
\textit{1 Alex Manoogian Street, 0025 Yerevan, Armenia } \vspace{0.5cm} \\
\textit{$^2$ International Centre for Theoretical Physics,} \\
\textit{Strada Costiera, 11 - 34014 Trieste, Italy}}
\maketitle

\begin{abstract}
We investigate the Wightman function, the vacuum expectation values of the
field squared and the energy-momentum tensor for a massless scalar field
with general curvature coupling parameter in spatially flat
Friedmann-Robertson-Walker universes with an arbitrary number of toroidally
compactified dimensions. The topological parts in the expectation values are
explicitly extracted and in this way the renormalization is reduced to that
for the model with trivial topology. In the limit when the comoving lengths
of the compact dimensions are very short compared to the Hubble length, the
topological parts coincide with those for a conformal coupling and they are
related to the corresponding quantities in the flat spacetime by standard
conformal transformation. This limit corresponds to the adiabatic
approximation. In the opposite limit of large comoving lengths of the
compact dimensions, in dependence of the curvature coupling parameter, two
regimes are realized with monotonic or oscillatory behavior of the vacuum
expectation values. In the monotonic regime and for nonconformally and
nonminimally coupled fields the vacuum stresses are isotropic and the
equation of state for the topological parts in the energy density and
pressures is of barotropic type. For conformal and minimal couplings the
leading terms in the corresponding asymptotic expansions vanish and the
vacuum stresses, in general, are anisotropic though the equation of state
remains of barotropic type. In the oscillatory regime, the amplitude of the
oscillations for the topological part in the expectation value of the field
squared can be either decreasing or increasing with time, whereas for the
energy-momentum tensor the oscillations are damping. The limits of validity
of the adiabatic approximation are discussed.
\end{abstract}

\bigskip

\section{Introduction}

Friedmann-Robertson-Walker (FRW) spacetimes are among the most popular
backgrounds in gravitational physics. There are several reasons for this.
First of all current observations give a strong motivation for the adoption
of the cosmological principle stating that at large scales the Universe is
homogeneous and isotropic and, hence, its large-scale structure is well
described by the FRW metric. Another motivation for the interest to FRW
geometries is related to the high symmetry of these backgrounds. Due this
symmetry numerous physical problems are exactly solvable and a better
understanding of physical effects in FRW models could serve as a handle to
deal with more complicated geometries. In particular, the investigation of
quantum effects at the early stages of the cosmological expansion has been a
subject of study in many research papers (see, for example, \cite%
{Birr82,Grib94} and references therein). The original motivations in this
research were mainly to avoid the initial singularity and to solve the
initial conditions problem for the dynamics of the Universe. The further
increase of the interest to this topic was related to the appearance of
inflationary cosmological scenarios \cite{Lind90}. During an inflationary
epoch, quantum fluctuations introduce inhomogeneities and may affect the
transition toward the true vacuum. These fluctuations provide a mechanism
which explains the origin of the primordial density perturbation needed to
explain the formation of the large-scale structure in the universe.

In many problems we need to consider the physical model on the background of
manifolds with compact spatial dimensions. In particular, the idea of
compact extra dimensions has been extensively used in supergravity and
superstring theories. From an inflationary point of view, universes with
compact dimensions, under certain conditions, should be considered a rule
rather than an exception \cite{Lind04}. The models of a compact universe
with non-trivial topology may play important roles by providing proper
initial conditions for inflation. The quantum creation of the universe
having toroidal spatial topology is discussed in \cite{Zeld84} and in
references \cite{Gonc85} within the framework of various supergravity
theories. In the case of non-trivial topology, the boundary conditions
imposed on quantum fields give rise to the modification of the spectrum for
vacuum fluctuations and, as a result, to the Casimir-type contributions in
the vacuum expectation values of physical observables (for the topological
Casimir effect and its role in cosmology see \cite{Most97,Appe87} and
references therein). In the Kaluza-Klein-type models, the topological
Casimir effect has been used as a stabilization mechanism for moduli fields
which parametrize the size and the shape of the extra dimensions. Recently,
the relevance of the Casimir energy as a model for dark energy needed for
the explanation of the present accelerated expansion of the universe has
been pointed out (see \cite{Milt03} and references therein).

In most work on the topological Casimir effect in cosmological backgrounds,
the results for the corresponding static counterparts were used replacing
the static length scales by comoving lengths in the cosmological bulk. This
procedure is valid in conformally invariant situations or under the
assumption of a quasi-adiabatic approximation. For non-conformal fields the
calculations should be done directly within the framework of quantum field
theory on time-dependent backgrounds. In a series of recent papers we have
considered the topological Casimir effect for a massive scalar field with
general curvature coupling \cite{Saha08} and for massive fermionic field
\cite{Saha08b} in background of the de Sitter spacetime with toroidally
compactified spatial dimensions. Due to the high symmetry of the de Sitter
spacetime in both cases the corresponding problems are exactly soluble.

In the present paper we will consider another exactly soluble problem for
the topological Casimir effect on background of a more general class of
spacetimes. Namely, we investigate the vacuum expectation values of the
field squared and the energy-momentum tensor for a massless scalar field
with arbitrary curvature coupling parameter induced by the compactness of
spatial dimensions in spatially flat FRW universes where the scale factor is
a power of the comoving time. The vacuum polarization and the particle
creation in the FRW cosmological models with trivial topology have been
considered in a large number of papers (see \cite{Birr82,Grib94,Most97} and
references therein for early research and \cite{Boya94}-\cite{Herd06} and
references therein for later developments). In particular, the vacuum
expectation values of the field squared and the energy-momentum tensor in
models with power law scale factors have been discussed in \cite%
{Bord97,Bord98},\cite{Ford77}-\cite{Beze96} (see also \cite{Grib94}). The
expectation value of the field squared is the crucial quantity in
discussions of the topological mass generation as a result of one-loop
quantum corrections and dynamical symmetry breaking. It also is of interest
in considerations of phase transitions in the early universe, in particular,
related to inflationary scenarios. In addition to describing the physical
structure of the quantum field at a given point, the expectation value of
the energy-momentum tensor acts as the source of gravity in the Einstein
equations. It therefore plays an important role in modelling a
self-consistent dynamics involving the gravitational field.

The present paper is organized as follows. In the next section we evaluate
the positive-frequency Wightman function in spatially flat FRW model with
topology $R^{p}\times (S^{1})^{q}$. By using the Abel-Plana summation
formula, we decompose this function into two parts: the first one is the
corresponding function in the geometry of the uncompactified FRW spacetime
and the second one is induced by the compactness of the spatial dimensions.
In section \ref{sec:vevphi2} we use the Wightman function for the evaluation
of the vacuum expectation value of the field squared. The asymptotic
behavior of the topological part is investigated in the early and late
stages of the cosmological evolution. In section \ref{sec:EMT} we consider
the vacuum expectation value of the energy-momentum tensor. The part in this
expectation value corresponding to the uncompactified FRW model is well
investigated in the literature and we are mainly concerned with the
topological part. The special case of the model with a single compact
dimension and the corresponding numerical results are discussed in section %
\ref{sec:Special}. The main results of the paper are summarized in Section %
\ref{sec:Conc}.

\section{Wightman function}

\label{sec:WF}

We consider a scalar field with curvature coupling parameter $\xi $\
evolving on background of the $(D+1)$-dimensional spatially flat FRW
spacetime with power law scale factor. Such a field is described by the
equation
\begin{equation}
\left( \nabla _{l}\nabla ^{l}+m^{2}+\xi R\right) \varphi =0,  \label{fieldeq}
\end{equation}%
where $\nabla _{l}$ denotes the covariant derivative and $R$ is the
curvature scalar of the background spacetime. For the most important special
cases of minimally and conformally coupled fields one has the values of the
curvature coupling parameter $\xi =0$ and $\xi =\xi _{D}\equiv (D-1)/4D$
respectively. The spatially flat FRW line element expressed in comoving time
coordinate is
\begin{equation}
ds^{2}=dt^{2}-a^{2}(t)\sum_{i=1}^{D}(dz^{i})^{2},\;a(t)=\alpha t^{c},
\label{ds2FRW}
\end{equation}%
with the curvature scalar
\begin{equation}
R=Dc\left[ (D+1)c-2\right] /t^{2}.  \label{R}
\end{equation}%
From the $(D+1)$-dimensional Einstein equations we find the energy density $%
\rho $ and the pressure $p$ for the source of the metric corresponding to (%
\ref{ds2FRW}):%
\begin{equation}
\rho =\frac{D(D-1)}{16\pi G}\frac{c^{2}}{t^{2}},\;p=\frac{D-1}{16\pi G}\frac{%
c(2-cD)}{t^{2}},  \label{rope}
\end{equation}%
where $G$ is the $(D+1)$-dimensional gravitational constant. The
corresponding equation of state has the form $p=[2/(cD)-1]\rho $. In the
special case%
\begin{equation}
c=2/(D+1),  \label{crad}
\end{equation}%
the equation of state is of the radiation type and the Ricci scalar
vanishes. The case $c=2/D$ corresponds to the dust-matter driven models.
Note that the line element (\ref{ds2FRW}) with $c>1$ describes the \ power
law \cite{Abbo87} and extended \cite{La89} inflationary models. Such a power
law expansion may be realized if a scalar field $\phi $ with an exponential
potential $V(\phi )=e^{-\lambda \phi }$ dominates the energy density of the
universe. The corresponding cosmological models admit power law inflationary
solutions of the form (\ref{ds2FRW}) (see, for instance, \cite{Burd88}) with
$c=4\lambda ^{-2}/(D-1)$. Solutions having power law scale factor with $%
c=2(n-1)(2n-1)/(D-2n)$ arise in higher-order gravity theories with the
Lagrangian $R^{n}$ \cite{Maed88} .

For the further discussion, in addition to the synchronous time coordinate $%
t $ it is convenient to introduce the conformal time $\tau $ in accordance
with $dt=a(t)d\tau $ or%
\begin{equation}
t=[\alpha (1-c)\tau ]^{1/(1-c)}.  \label{ttau}
\end{equation}%
Here we assume that $c\neq 1$. In the case $c=1$ one has $t=t_{0}e^{\alpha
\tau }$, with a constant $t_{0}$, and this case should be considered
separately. In terms of the conformal time the line element takes the form:%
\begin{equation}
ds^{2}=\Omega ^{2}(\tau )[d\tau ^{2}-\sum_{i=1}^{D}(dz^{i})^{2}],\;\Omega
(\tau )=\alpha \lbrack \alpha (1-c)\tau ]^{c/(1-c)},  \label{ds2Dd}
\end{equation}%
which is manifestly conformal to the Minkowski spacetime. Note that one has $%
0\leqslant \tau <\infty $ for $0<c<1$ and $-\infty <\tau \leqslant 0$ for $%
c>1$.

We will assume that the spatial coordinates $z^{l}$, $l=p+1,\ldots ,D$, are
compactified to $S^{1}$: $0\leqslant z^{l}\leqslant L_{l}$, and for the
other coordinates we have $-\infty \leqslant z^{l}\leqslant +\infty $, $%
l=1,\ldots ,p$. Hence, we consider the spatial topology $R^{p}\times
(S^{1})^{q}$ with $p+q=D$. For $p=0$, as a special case we obtain the
toroidally compactified FRW spacetime. Along the compact dimensions we will
consider the boundary conditions%
\begin{equation}
\varphi (\tau ,\mathbf{z}_{p},\mathbf{z}_{q}+\mathbf{e}_{l}L_{l})=e^{2\pi
i\alpha _{l}}\varphi (\tau ,\mathbf{z}_{p},\mathbf{z}_{q}),  \label{BC}
\end{equation}%
where $\mathbf{z}_{p}=(z^{1},\ldots ,z^{p})$, $\mathbf{z}_{q}=(z^{p+1},%
\ldots ,z^{D})$, and $\mathbf{e}_{l}$, $l=p+1,\ldots ,D$, is the unit vector
along the direction $z^{l}$. In (\ref{BC}) we have introduced phases with
constants $0\leqslant \alpha _{l}\leqslant 1$. The special cases $\alpha
_{l}=0$ and $\alpha _{l}=1/2$ correspond to periodic and antiperiodic
boundary conditions. The corresponding fields are also called as untwisted
and twisted ones.

Among the most important characteristics of the vacuum state are the vacuum
expectation values (VEVs) for the field squared and the energy-momentum
tensor. These VEVs are obtained from the Wightman function and its
derivatives in the coincidence limit of the arguments. For the topology
under consideration we will denote this function by $G_{p,q}^{+}(x,x^{\prime
})$. The Wightman function is also important in the consideration of the
response of particle detectors (see, for instance, \cite{Birr82}). Let $%
\left\{ \varphi _{\sigma }(x),\varphi _{\sigma }^{\ast }(x)\right\} $ be a
complete set of solutions to the classical field equation satisfying the
periodicity conditions (\ref{BC}) and $\sigma $ is a set of quantum numbers
specifying the solution. The Wightman function may be evaluated by using the
mode-sum formula
\begin{equation}
G_{p,q}^{+}(x,x^{\prime })=\langle 0|\varphi (x)\varphi (x^{\prime
})|0\rangle =\sum_{\sigma }\varphi _{\sigma }(x)\varphi _{\sigma }^{\ast
}(x^{\prime }),  \label{WF1}
\end{equation}%
where $|0\rangle $ is the amplitude for the vacuum state under consideration.

In accordance with the symmetry of the problem the mode solutions of the
field equation are separable: $\varphi _{\sigma }(x)=\phi (\tau )e^{i\mathbf{%
k}_{p}\cdot \mathbf{z}_{p}+i\mathbf{k}_{q}\cdot \mathbf{z}_{q}}$, where $%
\mathbf{k}_{p}=(k_{1},k_{2},\ldots ,k_{p})$ and $\mathbf{k}%
_{q}=(k_{p+1},k_{p+2},\ldots ,k_{D})$. For the components of the wave vector
along the uncompactified directions we have $-\infty <k_{l}<\infty $, $%
l=1,2,\ldots ,p$. The components along the compact dimensions are quantized
by the periodicity conditions (\ref{BC}) and for the corresponding
eigenvalues we have%
\begin{equation}
k_{l}=2\pi (n_{l}+\alpha _{l})/L_{l},\;n_{l}=0,\pm 1,\pm 2,\ldots
,\;l=p+1,\ldots ,D.  \label{kleig}
\end{equation}%
From the field equation (\ref{fieldeq}) we obtain an ordinary differential
equation for the function $\phi (\tau )$:
\begin{equation}
\frac{d^{2}\phi }{d\tau ^{2}}+\frac{(D-1)c}{(1-c)\tau }\frac{d\phi }{d\tau }%
+k^{2}\phi +\alpha ^{2}\left\{ [\alpha (1-c)\tau ]^{2c/(1-c)}m^{2}+\xi \frac{%
Dc\left[ (D+1)c-2\right] }{[\alpha (1-c)\tau ]^{2}}\right\} \phi =0,
\label{Eqphi}
\end{equation}%
where
\begin{equation}
k=\sqrt{\mathbf{k}_{p}^{2}+\mathbf{k}_{q}^{2}}.  \label{ktot}
\end{equation}%
For a massless scalar field, $m=0$, the general solution of this equation is
expressed in terms of a combination of Hankel functions:%
\begin{equation}
\phi (\tau )=\eta ^{b}\left[ c_{1}H_{\nu }^{(1)}(k\tau )+c_{2}H_{\nu
}^{(2)}(k\tau )\right] ,  \label{phisol}
\end{equation}%
with the notations%
\begin{equation}
\eta =|\tau |,\;b=\frac{cD-1}{2(c-1)},  \label{eta}
\end{equation}%
and%
\begin{equation}
\nu =\frac{1}{2|1-c|}\sqrt{(cD-1)^{2}-4\xi Dc\left[ (D+1)c-2\right] }.
\label{nu}
\end{equation}%
Note that $\nu $ is either real and nonnegative or pure imaginary. For a
conformally coupled field we have $\nu =1/2$ and for a minimally coupled
field $\nu =|b|$. Different choices of the coefficients in (\ref{phisol})
correspond to different vacuum states. We will consider the Bunch-Davies
vacuum \cite{Bunc78} for which $c_{1}=0$. With this choice, in the limit $%
c\rightarrow 0$ we recover the standard vacuum in the Minkowski spacetime
associated with the modes $k^{-1/2}\exp (i\mathbf{k}_{p}\cdot \mathbf{z}%
_{p}+i\mathbf{k}_{q}\cdot \mathbf{z}_{q}-ik\tau )$.

The corresponding eigenfunctions satisfying the periodicity conditions take
the form
\begin{equation}
\varphi _{\sigma }(x)=C_{\sigma }\eta ^{b}H_{\nu }^{(2)}(k\tau )e^{i\mathbf{k%
}_{p}\cdot \mathbf{z}_{p}+i\mathbf{k}_{q}\cdot \mathbf{z}_{q}}.
\label{eigfuncD}
\end{equation}%
The coefficient $C_{\sigma }$ with $\sigma =(\mathbf{k}_{p},n_{p+1},\ldots
,n_{D})$ is found from the orthonormalization condition
\begin{equation}
\int d^{D}x\sqrt{|g|}g^{00}\left[ \varphi _{\sigma }(x)\partial _{\tau
}\varphi _{\sigma ^{\prime }}^{\ast }(x)-\varphi _{\sigma ^{\prime }}^{\ast
}(x)\partial _{\tau }\varphi _{\sigma }(x)\right] =i\delta _{\sigma \sigma
^{\prime }},  \label{normcond}
\end{equation}%
where the integration goes over the spatial hypersurface $\tau =\mathrm{const%
}$, and $\delta _{\sigma \sigma ^{\prime }}$ is understood as the Kronecker
delta for discrete indices and as the Dirac delta-function for continuous
ones. By using the well known properties of the Hankel functions, this leads
to the result%
\begin{equation}
|C_{\sigma }|^{2}=\frac{[\alpha |1-c|]^{(D-1)c/(c-1)}}{2^{p+2}\pi
^{p-1}\alpha ^{D-1}V_{q}}e^{-(\nu -\nu ^{\ast })\pi i/2},  \label{normCD}
\end{equation}%
where and in what follows we denote by
\begin{equation}
V_{q}=L_{p+1}\cdots L_{D}  \label{Vq}
\end{equation}%
the volume of the compact subspace.

In the case $p=0$, $q=D$ (topology $(S^{1})^{D}$) and $\alpha _{l}=0$, $%
l=1,\ldots ,D$, there is also a normalizable zero mode with $\mathbf{k}%
_{D}=0 $ which is independent of the spatial coordinates (for the discussion
of zero modes in topologically nontrivial spaces see \cite{Ford89}). The
corresponding eigenfunction has the form $\varphi _{0}(\tau )=\eta
^{b}(c_{10}\eta ^{\nu }+c_{20}\eta ^{-\nu })$. From the normalization
condition we have the relation $c_{10}^{\ast }c_{20}-c_{10}c_{20}^{\ast }=i%
\mathrm{sign}(\tau )/(2\alpha \nu V_{D})$ for real $\nu $ and the relation $%
|c_{20}|^{2}-|c_{10}|^{2}=\mathrm{sign}(\tau )/(2|\nu |\alpha V_{D})$ for
imaginary $\nu $. These formulae give a relation between the coefficients $%
c_{10}$ and $c_{20}$ and one of them remains undetermined. This
arbitrariness arises because of freedom to choose the quantum state \cite%
{Ford89}. In the discussion below we will consider the contribution to the
VEVs from the nonzero modes. If the zero mode is present, its contribution
should be added separately.

Substituting the eigenfunctions (\ref{eigfuncD}) with the normalization
coefficient (\ref{normCD}) into the mode-sum formula for the Wightman
function, one finds
\begin{eqnarray}
G_{p,q}^{+}(x,x^{\prime }) &=&\frac{A(\eta \eta ^{\prime })^{b}}{2^{p}\pi
^{p+1}V_{q}}\int d\mathbf{k}_{p}\,e^{i\mathbf{k}_{p}\cdot \Delta \mathbf{z}%
_{p}}\sum_{\mathbf{n}_{q}\in \mathbf{Z}^{q}}e^{i\mathbf{k}_{q}\cdot \Delta
\mathbf{z}_{q}}  \notag \\
&&\times K_{\nu }(k\eta e^{\mathrm{sign}(\tau )\pi i/2})K_{\nu }(k\eta
^{\prime }e^{-\mathrm{sign}(\tau )\pi i/2}),  \label{WF2}
\end{eqnarray}%
where $\mathbf{n}_{q}=(n_{p+1},\ldots ,n_{D})$, $\Delta
z^{l}=z^{l}-z^{l\prime }$, and
\begin{equation}
A=\alpha ^{1-D}[\alpha |1-c|]^{(D-1)c/(c-1)}.  \label{A}
\end{equation}%
In formula (\ref{WF2}), for the further convenience, we have written the
Hankel function in terms of the modified Bessel function. For a scalar field
with periodic boundary conditions along compact dimensions ($\alpha _{l}=0$)
and for ${\mathrm{Re\,}}\nu \geqslant p/2$ the integral in (\ref{WF2}) has
an infrared divergence at $k_{p}=0$ coming from the modes $\mathbf{n}_{q}=0$%
. In this case the Bunch-Davies vacuum is well defined only if ${\mathrm{Re\,%
}}\nu <p/2$. For the case $p=0$ and $\alpha _{l}=0$, in (\ref{WF2}) $\mathbf{%
n}_{q}=\mathbf{n}_{D}\neq 0$ and the contribution of the additional zero
mode should be added.

We apply to the sum over $n_{p+1}$ in (\ref{WF2}) the Abel-Plana type
summation formula
\begin{eqnarray}
\sum_{n=-\infty }^{\infty }g(n+\beta )f(|n+\beta |) &=&\int_{0}^{\infty }du\,
\left[ g(u)+g(-u)\right] f(u)  \notag \\
&&+i\int_{0}^{\infty }du\left[ f(iu)-f(-iu)\right] \sum_{\lambda =\pm 1}%
\frac{g(i\lambda u)}{e^{2\pi (u+i\lambda \beta )}-1}.  \label{sumform}
\end{eqnarray}%
This formula is obtained by combining the summation formulae given in \cite%
{Beze08}. After the application of this formula, the term in the expression
of the Wightman function with the first integral on the right of (\ref%
{sumform}) corresponds to the Wightman function in the FRW model with $p+1$
uncompactified and $q-1$ toroidally compactified dimensions, $%
G_{p+1,q-1}^{+}(x,x^{\prime })$. As a result one finds the recurrence
formula
\begin{equation}
G_{p,q}^{+}(x,x^{\prime })=G_{p+1,q-1}^{+}(x,x^{\prime })+\Delta
_{p+1}G_{p,q}^{+}(x,x^{\prime }),  \label{G1decomp}
\end{equation}%
where the second term on the right is induced by the compactness of the $%
z^{p+1}$ - direction and is given by the expression%
\begin{eqnarray}
\Delta _{p+1}G_{p,q}^{+}(x,x^{\prime }) &=&\frac{A(\eta \eta ^{\prime })^{b}%
}{2(2\pi )^{p+1}V_{q-1}}\int d\mathbf{k}_{p}\,e^{i\mathbf{k}_{p}\cdot \Delta
\mathbf{z}_{p}}\sum_{\mathbf{n}_{q-1}\in \mathbf{Z}^{q-1}}e^{i\mathbf{k}%
_{q-1}\cdot \Delta \mathbf{z}_{q-1}}  \notag \\
&&\times \int_{0}^{\infty }dy\frac{y}{\sqrt{y^{2}+\mathbf{k}_{p}^{2}+\mathbf{%
k}_{\mathbf{n}_{q=1}}^{2}}}\sum_{\lambda =\pm 1}\frac{e^{-\lambda \sqrt{%
y^{2}+\mathbf{k}_{p}^{2}+\mathbf{k}_{\mathbf{n}_{q-1}}^{2}}\Delta z^{p+1}}}{%
e^{L_{p+1}\sqrt{y^{2}+\mathbf{k}_{p}^{2}+\mathbf{k}_{\mathbf{n}_{q-1}}^{2}}%
+2\pi i\lambda \alpha _{p+1}}-1}  \notag \\
&&\times \left\{ K_{\nu }(\eta y)\left[ I_{-\nu }(\eta ^{\prime }y)+I_{\nu
}(\eta ^{\prime }y)\right] +\left[ I_{\nu }(\eta y)+I_{-\nu }(\eta y)\right]
K_{\nu }(\eta ^{\prime }y)\right\} ,  \label{DeltGpq}
\end{eqnarray}%
where $I_{\nu }(z)$ is the modified Bessel function. Here the notations $%
V_{q-1}=L_{p+2},\ldots ,L_{D}$ and%
\begin{equation}
\mathbf{n}_{q-1}=(n_{p+2},\ldots ,n_{D})\text{, }\mathbf{k}_{\mathbf{n}%
_{q-1}}^{2}=\sum_{l=p+2}^{D}(2\pi n_{l}/L_{l})^{2},  \label{knD1+2}
\end{equation}%
are introduced. In formula (\ref{DeltGpq}) the integration with respect to
the angular part of $\mathbf{k}_{p}$ can be done explicitly. By using the
recurrence formula (\ref{G1decomp}), the Wightman function is presented in
the form%
\begin{equation}
G_{p,q}^{+}(x,x^{\prime })=G_{\mathrm{FRW}}^{+}(x,x^{\prime
})+\sum_{j=p}^{D-1}\Delta _{j+1}G_{j,D-j}^{+}(x,x^{\prime }),
\label{G1decomp1}
\end{equation}%
where $G_{\mathrm{FRW}}^{+}(x,x^{\prime })=G_{D,0}^{+}(x,x^{\prime })$ is
the corresponding function in the FRW model with spatial topology $R^{D}$
and the second term on the right is the topological part.

The integral representation for the function $G_{\mathrm{FRW}%
}^{+}(x,x^{\prime })$ is given by Eq. (\ref{WF2}) with $p=D$ and $q=0$. The
corresponding integral is explicitly evaluated by using the formula from
\cite{Prud86} and we find%
\begin{equation}
G_{\mathrm{FRW}}^{+}(x,x^{\prime })=\frac{A(\eta \eta ^{\prime })^{\frac{D-1%
}{2(c-1)}}}{2(2\pi )^{(D+1)/2}}\frac{\Gamma (D/2+\nu )\Gamma (D/2-\nu )}{%
(u^{2}-1)^{(D-1)/4}}P_{\nu -1/2}^{(1-D)/2}(u),  \label{GRW}
\end{equation}%
where $P_{\gamma }^{\mu }(u)$ is the associated Legendre function of the
first kind and
\begin{equation}
u=(|\Delta \mathbf{z}|^{2}-\eta ^{2}-\eta ^{\prime 2})/(2\eta \eta ^{\prime
}).  \label{u}
\end{equation}%
An alternative representation is obtained by making use of the relation
between the associated Legendre function and the hypergeometric function
(see, for instance, \cite{Abra76}):%
\begin{equation}
G_{\mathrm{FRW}}^{+}(x,x^{\prime })=\frac{A(\eta \eta ^{\prime })^{\frac{D-1%
}{2(c-1)}}}{(4\pi )^{(D+1)/2}}\frac{\Gamma (D/2+\nu )\Gamma (D/2-\nu )}{%
\Gamma ((D+1)/2)}F\left( \frac{D}{2}+\nu ,\frac{D}{2}-\nu ;\frac{D+1}{2},%
\frac{1-u}{2}\right) .  \label{GRW2}
\end{equation}%
Note that we have%
\begin{equation}
A(\eta \eta ^{\prime })^{\frac{D-1}{2(c-1)}}=|1-c|^{D-1}(tt^{\prime
})^{(1-D)/2}.  \label{Aeta}
\end{equation}%
In the special case $D=3$ and $\xi =0$ the formula (\ref{GRW2}) coincides
with the result given in \cite{Bunc78} (with the missprint in the
coefficient corrected, see also \cite{Birr82}). Shifting the time coordinate
$t\rightarrow t+ct_{0}$, with a constant $t_{0}$ and taking the limit $%
c\rightarrow \infty $ for fixed $t$ and $t_{0}$, from (\ref{ds2FRW}) we
obtain the corresponding line element in de Sitter spacetime with the scale
factor $a(t)=e^{t/t_{0}}$. In this limit the expression (\ref{Aeta}) tends
to $t_{0}^{1-D}$ and from (\ref{GRW2}) we recover the expression for the
Wightman function in $(D+1)$-dimensional de Sitter spacetime.

\section{VEV of the field squared}

\label{sec:vevphi2}

The VEV of the field squared in FRW model with topology $R^{p}\times
(S^{1})^{q}$ is obtained from the two-point function $G_{p,q}^{+}(x,x^{%
\prime })$ taking the coincidence limit of the arguments. In this limit the
Wightman function diverges and a renormalization procedure is necessary. The
crucial is that under the toroidal compactification the local geometry is
not changed and the divergences are contained in the part corresponding to
the uncompactified FRW spacetime. As we have already extracted this part,
the renormalization is reduced to that in the uncompactified FRW model which
is well investigated in the literature (see for example \cite{Birr82, Grib94}
and references given therein). For the VEV of the field squared we have the
recurrence formula
\begin{equation}
\langle \varphi ^{2}\rangle _{p,q}=\langle \varphi ^{2}\rangle
_{p+1,q-1}+\Delta _{p+1}\langle \varphi ^{2}\rangle _{p,q},
\label{phi2decomp}
\end{equation}%
where $\langle \varphi ^{2}\rangle _{p+1,q-1}$ is the expectation value in
the FRW spacetime with the spatial topology $R^{p+1}\times (S^{1})^{q-1}$
and the second term on the right is the part due to the compactness of the $%
z^{p+1}$ - direction. The latter is directly obtained from (\ref{DeltGpq})
taking the coincidence limit of the arguments. The integral over $\mathbf{k}%
_{p}$ is evaluated by making use of the formula%
\begin{equation}
\sum_{\lambda =\pm 1}\int d\mathbf{k}_{p}\,\frac{(\mathbf{k}%
_{p}^{2}+a^{2})^{-1/2}}{e^{L_{p+1}\sqrt{\mathbf{k}_{p}^{2}+a^{2}}+2\pi
i\lambda \alpha _{p+1}}-1}=4(2\pi )^{(p-1)/2}\sum_{n=1}^{\infty }\frac{\cos
(2\pi n\alpha _{p+1})}{(nL_{p+1})^{p-1}}f_{(p-1)/2}(nL_{p+1}a),  \label{rel1}
\end{equation}%
with the notation $f_{\nu }(x)=x^{\nu }K_{\nu }(x)$. Now for the topological
part in (\ref{phi2decomp}) we find%
\begin{eqnarray}
\Delta _{p+1}\langle \varphi ^{2}\rangle _{p,q} &=&\frac{4A\eta ^{2b}}{(2\pi
)^{(p+3)/2}V_{q-1}}\sum_{\mathbf{n}_{q-1}\in \mathbf{Z}^{q-1}}\int_{0}^{%
\infty }dyy\left[ I_{-\nu }(y\eta )+I_{\nu }(y\eta )\right] K_{\nu }(y\eta )
\notag \\
&&\times \sum_{n=1}^{\infty }\frac{\cos (2\pi n\alpha _{p+1})}{%
(nL_{p+1})^{p-1}}f_{(p-1)/2}(nL_{p+1}\sqrt{y^{2}+\mathbf{k}_{\mathbf{n}%
_{q}-1}^{2}}).  \label{phi2}
\end{eqnarray}%
This integral representation is valid for ${\mathrm{Re\,}}\nu <1$.

Similar to (\ref{G1decomp1}), we find the following decomposition for the
VEV of the field squared:%
\begin{equation}
\langle \varphi ^{2}\rangle _{p,q}=\langle \varphi ^{2}\rangle _{\mathrm{FRW}%
}+\langle \varphi ^{2}\rangle _{p,q}^{\mathrm{(t)}},\;\langle \varphi
^{2}\rangle _{p,q}^{\mathrm{(t)}}=\sum_{j=p}^{D-1}\Delta _{j+1}\langle
\varphi ^{2}\rangle _{j,D-j},  \label{phi2decomp2}
\end{equation}%
where $\langle \varphi ^{2}\rangle _{\mathrm{FRW}}=\langle \varphi
^{2}\rangle _{D,0}$ is the VEV in the spatial topology $R^{D}$ and the part $%
\langle \varphi ^{2}\rangle _{p,q}^{\mathrm{(t)}}$ is induced by the
nontrivial topology. As it can be seen from formula (\ref{phi2}), the
topological part has the following general structure: $\langle \varphi
^{2}\rangle _{p,q}^{\mathrm{(t)}}=t^{1-D}f(L_{p+1}/\eta ,\ldots ,L_{D}/\eta
) $, where the form of the function $f(x_{1},\ldots ,x_{q})$ is directly
obtained from \ (\ref{phi2}) and (\ref{phi2decomp2}). The ratios in the
arguments of this function may also be written as
\begin{equation}
L_{l}/\eta =(|1-c|/c)L_{l}^{\mathrm{(c)}}/r_{\mathrm{H}},  \label{LcrH}
\end{equation}%
with $L_{l}^{\mathrm{(c)}}=\Omega L_{l}$ and $r_{\mathrm{H}}=t/c$ being the
comoving length of the compact dimension and the Hubble length respectively.

For a conformally coupled massless scalar field one has $\nu =1/2$ and $%
\left[ I_{-\nu }(x)+I_{\nu }(x)\right] K_{\nu }(x)=1/x$. The integral in (%
\ref{phi2}) is explicitly evaluated and we find%
\begin{equation}
\Delta _{p+1}\langle \varphi ^{2}\rangle _{p,q}=\frac{2A\eta ^{2b-1}}{(2\pi
)^{p/2+1}V_{q-1}}\sum_{\mathbf{n}_{q-1}\in \mathbf{Z}^{q-1}}\sum_{n=1}^{%
\infty }\frac{\cos (2\pi n\alpha _{p+1})}{(nL_{p+1})^{p}}f_{p/2}(nL_{p+1}k_{%
\mathbf{n}_{q}-1}).  \label{phi21}
\end{equation}%
We could obtain this result directly from the corresponding formula in the
Minkowski spacetime with topology $R^{p}\times (S^{1})^{q}$, by taking into
account that $A\eta ^{2b-1}=\Omega ^{1-D}$ and by using the fact that two
problems are conformally related: $\langle \varphi ^{2}\rangle _{p,q}^{%
\mathrm{(t)}}=\Omega ^{1-D}\langle \varphi ^{2}\rangle _{p,q}^{\mathrm{(t,M)}%
}$. The expression for $\langle \varphi ^{2}\rangle _{p,q}^{\mathrm{(t,M)}}$
is directly obtained from (\ref{phi21}) and is valid for arbitrary values of
the curvature coupling parameter $\xi $. Note that we have $\nu =1/2$ in the
general case of the curvature coupling parameter for the power law expansion
with $c=2/(D+1)$ which corresponds to radiation driven models. In this case
the topological part of the VEV is given by formula (\ref{phi21}) with $%
b=-1/2$ and $A=\alpha ^{-D-1}(D+1)^{2}/(D-1)^{2}$.

Now we turn to the investigation of the topological part $\Delta
_{p+1}\langle \varphi ^{2}\rangle _{p,q}$ in the VEV of the field squared in
the asymptotic regions of the ratio $L_{p+1}/\eta $. For small values of
this ratio, $L_{p+1}/\eta \ll 1$, we introduce the new integration variable $%
y=L_{p+1}x$. By taking into account that for large values $x$ one has $\left[
I_{-\nu }(x)+I_{\nu }(x)\right] K_{\nu }(x)\approx 1/x$, we find that to the
leading order $\Delta _{p+1}\langle \varphi ^{2}\rangle _{p,q}$ coincides
with the corresponding result for a conformally coupled massless field:%
\begin{equation}
\Delta _{p+1}\langle \varphi ^{2}\rangle _{p,q}\approx \frac{2\Omega ^{1-D}}{%
(2\pi )^{p/2+1}V_{q-1}}\sum_{\mathbf{n}_{q-1}\in \mathbf{Z}%
^{q-1}}\sum_{n=1}^{\infty }\frac{\cos (2\pi n\alpha _{p+1})}{(nL_{p+1})^{p}}%
f_{p/2}(nL_{p+1}k_{\mathbf{n}_{q}-1}),\;L_{p+1}/\eta \ll 1.
\label{phi2small}
\end{equation}%
In terms of the comoving time coordinate the topological part behaves as $%
\Delta _{p+1}\langle \varphi ^{2}\rangle _{p,q}\propto t^{(1-D)c}$. The
limit under consideration corresponds to early stages of the cosmological
expansion ($t\rightarrow 0$) for $c>1$ and to late stages ($t\rightarrow
\infty $) for $0<c<1$. Note that, by taking into account (\ref{LcrH}), the
condition $L_{p+1}/\eta \ll 1$ can also be written in the form $L_{p+1}^{%
\mathrm{(c)}}\ll r_{\mathrm{H}}$. Hence, the asymptotic formula (\ref%
{phi2small}) corresponds to small proper lengths of compact dimensions
compared with the Hubble length. The expression on the right hand side of (%
\ref{phi2small}) is obtained from the corresponding expression in Minkowski
spacetime with the topology $R^{p}\times (S^{1})^{q}$ and with the lengths
of the compact dimensions $L_{p+1},\ldots ,L_{D}$, replacing $L_{l}$ by the
comoving lengths $L_{l}^{\mathrm{(c)}}$. Hence, (\ref{phi2small})
corresponds to the adiabatic approximation. In this regime the topological
part monotonically decreases with increasing scale factor.

In the opposite limit of large values for the ratio $L_{p+1}/\eta $, the
behavior of the topological part is qualitatively different for real and
imaginary values of the parameter $\nu $. For positive values $\nu $, by
using the asymptotic formulae for the modified Bessel functions for small
values of the arguments, to the leading order we find%
\begin{eqnarray}
\Delta _{p+1}\langle \varphi ^{2}\rangle _{p,q} &\approx &\frac{2^{\nu
+1}A\eta ^{2b-2\nu }\Gamma (\nu )}{(2\pi )^{(p+3)/2}V_{q-1}}\sum_{\mathbf{n}%
_{q-1}\in \mathbf{Z}^{q-1}}\sum_{n=1}^{\infty }\frac{\cos (2\pi n\alpha
_{p+1})}{(nL_{p+1})^{p+1-2\nu }}  \notag \\
&&\times f_{(p+1)/2-\nu }(nL_{p+1}k_{\mathbf{n}_{q}-1}),\;L_{p+1}/\eta \gg 1.
\label{phi2large}
\end{eqnarray}%
In terms of the synchronous time coordinate we have $\Delta _{p+1}\langle
\varphi ^{2}\rangle _{p,q}\propto t^{2(c-1)\nu -cD+1}$. This limit
corresponds to early stages of the cosmological expansion ($t\rightarrow 0$)
for $0<c<1$ and to late stages of the cosmological expansion ($t\rightarrow
\infty $) when $c>1$. Note that for $0<c<1/D$ and $\xi <0$ one has $b>\nu $
and the topological part in the expectation value of $\varphi ^{2}$
increases with increasing comoving time $t$. For a minimally coupled scalar
field and for $c\in (0,1/D)\cup (1,\infty )$ one has $\nu =b$, and in the
limit under consideration the VEV\ tends to finite nonzero value. In the
case $\nu =0$ the corresponding asymptotic formula is obtained from (\ref%
{phi2large}) replacing $\Gamma (\nu )\rightarrow 4\ln (L_{p+1}/\eta )$ and
after substituting $\nu =0$. The limit under consideration corresponds to
large comoving length of the compact dimension compared to the Hubble
length: $L_{p+1}^{\mathrm{(c)}}\gg r_{\mathrm{H}}$.

For pure imaginary values $\nu $ and in the same limit $L_{p+1}/\eta \gg 1$,
we use the formula%
\begin{equation}
K_{\nu }(z)\left[ I_{-\nu }(z)+I_{\nu }(z)\right] \approx {\mathrm{Re}}\left[
\frac{\Gamma (\nu )(z/2)^{-2\nu }}{\Gamma (1-\nu )}\right] ,  \label{KIapp}
\end{equation}%
valid for small values $z$. Now we can see that in the leading order%
\begin{equation}
\Delta _{p+1}\langle \varphi ^{2}\rangle _{p,q}\approx 4AB\eta ^{2b}\frac{%
\cos [2|\nu |\ln (L_{p+1}/\eta )+\phi ]}{(2\pi )^{(p+3)/2}L_{p+1}^{p}V_{q}}%
,\;L_{p+1}/\eta \gg 1,  \label{phi2largeim}
\end{equation}%
where the constants $B$ and $\phi $ are defined by the relation
\begin{equation}
Be^{i\phi }=2^{i|\nu |}\Gamma (i|\nu |)\sum_{n=1}^{\infty }\frac{\cos (2\pi
n\alpha _{p+1})}{n^{p+1-2i|\nu |}}\sum_{\mathbf{n}_{q-1}\in \mathbf{Z}%
^{q-1}}f_{(p+1)/2-i|\nu |}(nL_{p+1}k_{\mathbf{n}_{q}-1}).  \label{Bphi}
\end{equation}%
Hence, the behaviour of the topological part in the VEV for the field
squared is oscillatory in the early (late) stages of the cosmological
expansion for $0<c<1$ ($c>1$). The amplitude of these oscillations behaves
like $t^{1-cD}$ and the distance between the nearest zeros linearly
increases with increasing $t$. Note that for $0<c<1/D$ the oscillation
amplitude increases with time.

From the asymptotic analysis given above it follows that in the special case
of a minimally coupled field the topological part $\Delta _{p+1}\langle
\varphi ^{2}\rangle _{p,q}$ tends to finite nonzero value at early stages of
the cosmological expansion when $0<c<1/D$. At late (early) stages and for $%
c<1$ ($c>1$) the topological part behaves like $t^{(1-D)c}$. At early times
and for $1/D<c<1$ one has $\Delta _{p+1}\langle \varphi ^{2}\rangle
_{p,q}\propto t^{2(1-cD)}$.

\section{VEV of the energy-momentum tensor}

\label{sec:EMT}

Having the Wightman function and the VEV\ of the field squared, the
expectation value of the energy-momentum tensor is evaluated with the help
of the formula
\begin{equation}
\langle T_{ik}\rangle _{p,q}=\lim_{x^{\prime }\rightarrow x}\partial
_{i}\partial _{k}^{\prime }G_{p,q}^{+}(x,x^{\prime })+\left[ \left( \xi -%
\frac{1}{4}\right) g_{ik}\nabla _{l}\nabla ^{l}-\xi \nabla _{i}\nabla
_{k}-\xi R_{ik}\right] \langle \varphi ^{2}\rangle _{p,q},  \label{emtvev1}
\end{equation}%
where $R_{ik}$ is the Ricci tensor for the spatially flat FRW spacetime with
the components
\begin{equation}
R_{00}=\frac{Dc\tau ^{-2}}{c-1},\;R_{ik}=-\delta _{ik}\frac{c(Dc-1)}{%
(1-c)^{2}\tau ^{2}},\;i,k=1,\ldots ,D.  \label{Rik}
\end{equation}%
In Eq. (\ref{emtvev1}) we have used the expression of the metric
energy-momentum tensor for a scalar field which differs from the standard
one (see, for example, \cite{Birr82}) by the term vanishing on the solutions
of the field equation \cite{Saha04}. The topological part in the VEV of the
energy-momentum tensor is obtained substituting (\ref{G1decomp}) and (\ref%
{phi2decomp}) into formula (\ref{emtvev1}). In this calculation we need the
covariant d'Alembertian acted on the topological part of the field squared:%
\begin{eqnarray}
\nabla _{i}\nabla ^{i}(\Delta _{p+1}\langle \varphi ^{2}\rangle _{p,q}) &=&%
\frac{2^{(1-p)/2}Ac}{\pi ^{(p+3)/2}V_{q-1}(c-1)}\sum_{\mathbf{n}_{q-1}\in
\mathbf{Z}^{q-1}}\sum_{n=1}^{\infty }\frac{\cos (2\pi n\alpha _{p+1})}{%
(nL_{p+1})^{p-1}}\int_{0}^{\infty }dy\,y^{3-2b}  \notag \\
&&\times f_{(p-1)/2}(nL_{p+1}\sqrt{y^{2}+\mathbf{k}_{\mathbf{n}_{q}-1}^{2}}%
)\left( \frac{1}{z}\partial _{z}\right) \widetilde{I}_{\nu }(z)\widetilde{K}%
_{\nu }(z)|_{z=\eta y}.  \label{Dalamphi}
\end{eqnarray}%
Here and in the discussion below, to simplify our notations, we have
introduced the functions
\begin{equation}
\widetilde{K}_{\nu }(z)=z^{b}K_{\nu }(z),\;\widetilde{I}_{\nu }(z)=z^{b}%
\left[ I_{\nu }(z)+I_{-\nu }(z)\right] ,  \label{KItilde}
\end{equation}%
and the function $f_{\nu }(x)$ is defined after formula (\ref{rel1}).

For the VEV of the energy-momentum tensor we have the following recurrence
formula%
\begin{equation}
\langle T_{l}^{k}\rangle _{p,q}=\langle T_{l}^{k}\rangle _{p+1,q-1}+\Delta
_{p+1}\langle T_{l}^{k}\rangle _{p,q},  \label{EMTdecomp}
\end{equation}%
where the second term on the right hand side is due to the compactness of
the direction $z^{p+1}$. For the topological part in the energy density we
find%
\begin{eqnarray}
\Delta _{p+1}\langle T_{0}^{0}\rangle _{p,q} &=&\frac{4A\Omega ^{-2}}{(2\pi
)^{(p+3)/2}V_{q-1}}\sum_{n=1}^{\infty }\frac{\cos (2\pi n\alpha _{p+1})}{%
(nL_{p+1})^{p-1}}\int_{0}^{\infty }dy\,y^{3-2b}  \notag \\
&&\times \sum_{\mathbf{n}_{q-1}\in \mathbf{Z}^{q-1}}f_{(p-1)/2}(nL_{p+1}%
\sqrt{y^{2}+\mathbf{k}_{\mathbf{n}_{q}-1}^{2}})F^{(0)}(\eta y),  \label{T00}
\end{eqnarray}%
with the notation
\begin{equation}
F^{(0)}(z)=\frac{1}{2}\widetilde{I}_{\nu }^{\prime }(z)\widetilde{K}_{\nu
}^{\prime }(z)+\frac{D\xi c}{z(1-c)}(\widetilde{I}_{\nu }\widetilde{K}_{\nu
})^{\prime }-\frac{1}{2}\left[ 1-\frac{\xi D(D-1)c^{2}}{z^{2}(1-c)^{2}}%
\right] \widetilde{I}_{\nu }\widetilde{K}_{\nu }.  \label{F0}
\end{equation}%
In a similar way, for the vacuum stresses the following representation holds
(no summation over $i$)%
\begin{eqnarray}
\Delta _{p+1}\langle T_{i}^{i}\rangle _{p,q} &=&\frac{4A\Omega ^{-2}}{(2\pi
)^{(p+3)/2}V_{q-1}}\sum_{\mathbf{n}_{q-1}\in \mathbf{Z}^{q-1}}\sum_{n=1}^{%
\infty }\frac{\cos (2\pi n\alpha _{p+1})}{(nL_{p+1})^{p-1}}\int_{0}^{\infty
}dy\,y^{3-2b}  \notag \\
&&\times \left[ f_{(p-1)/2}(z)F(\eta y)-f_{p}^{(i)}(z)\frac{\widetilde{I}%
_{\nu }(\eta y)\widetilde{K}_{\nu }(\eta y)}{(nL_{p+1}y)^{2}}\right]
_{z=nL_{p+1}\sqrt{y^{2}+\mathbf{k}_{\mathbf{n}_{q}-1}^{2}}},  \label{Tiin}
\end{eqnarray}%
where we have used the notations%
\begin{eqnarray}
f_{p}^{(i)}(y) &=&f_{(p+1)/2}(y),\;i=1,2,\ldots p,  \notag \\
f_{p}^{(p+1)}(y) &=&-pf_{(p+1)/2}(y)-y^{2}f_{(p-1)/2}(y),  \label{fp} \\
f_{p}^{(i)}(y) &=&(nL_{p+1}k_{i})^{2}f_{(p-1)/2}(y),\;i=p+2,\ldots D.  \notag
\end{eqnarray}%
In formula (\ref{Tiin}), the function $F(z)$ is defined as%
\begin{eqnarray}
F(z) &=&2\left( \xi -\frac{1}{4}\right) \widetilde{I}_{\nu }^{\prime }%
\widetilde{K}_{\nu }^{\prime }-\frac{1}{z}\frac{c\xi }{1-c}(\widetilde{I}%
_{\nu }\widetilde{K}_{\nu })^{\prime }  \notag \\
&&+2\left[ \xi -\frac{1}{4}-Dc\xi \frac{(Dc-2)(\xi -\xi _{D})+\xi c}{%
z^{2}(1-c)^{2}}\right] \widetilde{I}_{\nu }(z)\widetilde{K}_{\nu }(z).
\label{Fz}
\end{eqnarray}%
Now it can be checked that the topological parts in the VEVs of the field
squared and the energy-momentum tensor obey the trace relation
\begin{equation}
\Delta _{p+1}\langle T_{i}^{i}\rangle _{p,q}=D(\xi -\xi _{D})\nabla
_{i}\nabla ^{i}(\Delta _{p+1}\langle \varphi ^{2}\rangle _{p,q}).
\label{TraceRel}
\end{equation}%
In particular, for a conformally coupled field the corresponding
energy-momentum tensor is traceless.

The recurrence relation (\ref{EMTdecomp}) allows to present the VEV of the
energy-momentum tensor as the sum%
\begin{equation}
\langle T_{i}^{k}\rangle _{p,q}=\langle T_{i}^{k}\rangle _{\mathrm{FRW}%
}+\langle T_{i}^{k}\rangle _{p,q}^{\mathrm{(t)}},\;\langle T_{i}^{k}\rangle
_{p,q}^{\mathrm{(t)}}=\sum_{j=p}^{D-1}\Delta _{j+1}\langle T_{i}^{k}\rangle
_{j,D-j},  \label{TikDecomp}
\end{equation}%
where $\langle T_{i}^{k}\rangle _{\mathrm{FRW}}$ is the part corresponding
to the uncompactified FRW spacetime and $\langle T_{i}^{k}\rangle _{p,q}^{%
\mathrm{(t)}}$ is induced by the the nontrivial topology. The first term is
well investigated in the \ literature (see\ \cite{Birr82,Grib94}\ and
references therein) and in the following we will concentrate on the
topological part. Note that this part has the general structure $\langle
T_{i}^{k}\rangle _{p,q}^{\mathrm{(t)}}=t^{-D-1}\delta
_{i}^{k}f^{(i)}(L_{p+1}/\eta ,\ldots ,L_{D}/\eta )$, where the form of the
functions $f^{(i)}(x_{1},\ldots ,x_{q})$ directly follows from formulae (\ref%
{T00}) and (\ref{Tiin}). Recall that the ratios in the arguments of these
functions may also be written in the form (\ref{LcrH}).

The expressions for the VEV of the energy-momentum tensor are further
simplified for a conformally coupled scalar field. In this case, after some
calculations we find (no summation over $i$):%
\begin{equation}
\Delta _{p+1}\langle T_{i}^{i}\rangle _{p,q}=-\frac{2\Omega ^{-D-1}}{(2\pi
)^{p/2+1}V_{q-1}}\sum_{n=1}^{\infty }\sum_{\mathbf{n}_{q-1}\in \mathbf{Z}%
^{q-1}}\frac{\cos (2\pi n\alpha _{p+1})}{(nL_{p+1})^{p+2}}f_{\mathrm{(c)}%
p}^{(i)}(nL_{p+1}k_{\mathbf{n}_{q}-1}),  \label{TiiConf}
\end{equation}%
where%
\begin{eqnarray}
f_{\mathrm{(c)}p}^{(i)}(x) &=&f_{p/2+1}(x),\;i=0,1,\ldots ,p,  \notag \\
f_{\mathrm{(c)}p}^{(p+1)}(x) &=&-(p+1)f_{p/2+1}(x)-x^{2}f_{p/2}(x),
\label{fconf} \\
f_{\mathrm{(c)}p}^{(i)}(x) &=&(nL_{p+1}k_{i})^{2}f_{p/2}(x),\;i=p+2,\ldots
,D.  \notag
\end{eqnarray}%
This result is also obtained from the corresponding formulae in the
toroidally compactified Minkowski spacetime by using the standard relation
for the VEVs in conformally related problems. Formula (\ref{TiiConf}) for a
conformally coupled field holds for any conformally flat bulk with general
scale factor $\Omega (\tau )$. As wee see, for a conformally coupled field
the stresses along the uncompactified dimensions are equal to the vacuum
energy density.

In the special case of power law expansion with $c=2/(D+2)$ and for general $%
\xi $ we again have $\nu =1/2$. After the integration over $y$, in this case
we find%
\begin{eqnarray}
\Delta _{p+1}\langle T_{i}^{i}\rangle _{p,q} &=&\frac{2\Omega ^{-D-1}}{(2\pi
)^{p/2+1}V_{q-1}}\sum_{\mathbf{n}_{q-1}\in \mathbf{Z}^{q-1}}\sum_{n=1}^{%
\infty }\frac{\cos (2\pi n\alpha _{p+1})}{(nL_{p+1})^{p+2}}  \notag \\
&&\times \left[ 2f_{i}\frac{\xi -\xi _{D}}{D-1}\left( \frac{nL_{p+1}}{\eta }%
\right) ^{2}f_{p/2}(nL_{p+1}k_{\mathbf{n}_{q}-1})-f_{\mathrm{(c)}%
p}^{(i)}(nL_{p+1}k_{\mathbf{n}_{q}-1})\right] ,  \label{TiiSpc}
\end{eqnarray}%
where $f_{0}=D$ and $f_{i}=1$ for $i=1,2,\ldots ,D$. Recall that this
special case corresponds to the radiation driven expansion and the
corresponding Ricci scalar is zero.

The formulae given above for the VEV of the energy-momentum tensor in the
general case of the curvature coupling are simplified in the asymptotic
regions of the ratio $L_{p+1}/\eta $. For small values of this ratio the
arguments of the modified Bessel functions in the expressions for the VEV of
the energy-momentum tensor are large. In the leading order we have $%
F^{(0)}(z)\approx -z^{2b-1}$ and to this order the first term in the square
brackets in the expression (\ref{Tiin}) for the vacuum stresses does no
contribute. The integral over $y$ is explicitly evaluated and we find%
\begin{equation}
\Delta _{p+1}\langle T_{i}^{i}\rangle _{p,q}\approx -\frac{2\Omega ^{-D-1}}{%
(2\pi )^{p/2+1}V_{q-1}}\sum_{n=1}^{\infty }\sum_{\mathbf{n}_{q-1}\in \mathbf{%
Z}^{q-1}}\frac{\cos (2\pi n\alpha _{p+1})}{(nL_{p+1})^{p+2}}f_{\mathrm{(c)}%
p}^{(i)}(nL_{p+1}k_{\mathbf{n}_{q}-1}),\;L_{p+1}/\eta \ll 1.
\label{Tiismall}
\end{equation}%
As we see, in the leading order the VEV coincides with the corresponding
expression for a conformally coupled field. This limit corresponds to late
stages of the cosmological expansion ($t\rightarrow \infty $) for $0<c<1$
and to early stages ($t\rightarrow 0$) for $c>1$. In terms of the comoving
time coordinate, the topological part behaves as $\Delta _{p+1}\langle
T_{i}^{i}\rangle _{p,q}\propto t^{-(D+1)c}$. By using the relation (\ref%
{LcrH}), we see that the asymptotic expression (\ref{Tiismall}) corresponds
to small proper length of the compact dimension compared to the Hubble
length.

For large values of the ratio $L_{p+1}/\eta $ and in the case of positive $%
\nu $ for the topological part in the VEV of the energy-momentum tensor one
has (no summation over $i$)%
\begin{equation}
\Delta _{p+1}\langle T_{i}^{i}\rangle _{p,q}\approx \frac{F^{(i)}}{(\Omega
\eta )^{2}}\Delta _{p+1}\langle \varphi ^{2}\rangle _{p,q},\;L_{p+1}/\eta
\gg 1,  \label{T00largeL}
\end{equation}%
where the asymptotic expression for the field squared is given by Eq. (\ref%
{phi2large}) and we have defined%
\begin{eqnarray}
F^{(0)} &=&\frac{1}{2}(b-\nu )^{2}+2D\xi c\frac{b-\nu }{1-c}+\frac{\xi
D(D-1)c^{2}}{2(1-c)^{2}},  \notag \\
F^{(l)} &=&2\left( \xi -\frac{1}{4}\right) (b-\nu )^{2}-2c\xi \frac{b-\nu }{%
1-c}-2Dc\xi \frac{(Dc-2)(\xi -\xi _{D})+\xi c}{(1-c)^{2}},  \label{F0l}
\end{eqnarray}%
with $l=1,\ldots ,D$. As it is seen, the leading terms in the vacuum
stresses are isotropic. In terms of the comoving time coordinate, we have
the behaviour $\Delta _{p+1}\langle T_{i}^{i}\rangle _{p,q}\propto
t^{2(c-1)\nu -cD-1}$. Note that this holds for the total topological part $%
\langle T_{i}^{i}\rangle _{p,q}^{\mathrm{(t)}}$ as well. The limit under
consideration corresponds to late times of the cosmological expansion ($%
t\rightarrow \infty $) for $c>1$ and to early times ($t\rightarrow 0$) for $%
0<c<1$, or alternatively, to large comoving length of the compact dimension
in units of the Hubble length: $L_{p+1}^{\mathrm{(c)}}/r_{\mathrm{H}}\gg 1$.
In the case $\nu =0$ the asymptotics of the topological part in the VEV of
the energy-momentum tensor are still given by relation (\ref{T00largeL}),
where now%
\begin{equation}
F^{(0)}=\frac{\xi D(D-1)}{(1-1/c)^{2}},\;F^{(l)}=-\frac{\xi }{2}\left[ 1+%
\frac{D-1}{(1-1/c)^{2}}\right] ,\;\nu =0.  \label{Finu0}
\end{equation}%
Note that the coefficient in (\ref{T00largeL}) does not depend on $p$ and,
hence, similar relation takes place between the total topological parts in
the VEVs of the field squared and the energy-momentum tensor (given by
formulae (\ref{phi2decomp2}) and (\ref{TikDecomp})). In the limit under
consideration the vacuum stresses are isotropic and the equation of state
for the topological parts in the vacuum energy density and pressures is of
the barotropic type (no summation over $i$): $\langle T_{i}^{i}\rangle
_{p,q}^{\mathrm{(t)}}\approx (F^{(1)}/F^{(0)})\langle T_{0}^{0}\rangle
_{p,q}^{\mathrm{(t)}}$, $i=1,2,\ldots ,D$. In the case of a minimally
coupled scalar field and for $1/D<c<1$ we have $\nu =-b$ and, as it follows
from (\ref{F0l}), $\Delta _{p+1}\langle T_{i}^{i}\rangle _{p,q}\approx
-\Delta _{p+1}\langle T_{0}^{0}\rangle _{p,q}$, $i=1,2,\ldots ,D$ (no
summation over $i$). This corresponds to the equation of state for stiff
fluid with pressure equal to the energy density.

For a minimally coupled scalar field in the case $c\in (0,1/D)\cup (1,\infty
)$ (in this case $\nu =b$) and for a conformally coupled field the functions
$F^{(i)}$ in (\ref{F0l}) vanish. In these cases we should keep the next
terms in the corresponding asymptotic expansions. For a conformally coupled
field the behavior is described by the exact formula (\ref{TiiConf}). In the
case of a minimally coupled field and for $c\in (0,1/D)\cup (1,\infty )$, to
the leading order we have (no summation over $i$):%
\begin{eqnarray}
\frac{\Delta _{p+1}\langle T_{0}^{0}\rangle _{p,q}}{b-1} &\approx &\frac{%
\Delta _{p+1}\langle T_{i}^{i}\rangle _{p,q}}{b-2}\approx \frac{%
2^{b+1}\Gamma (b)A\Omega ^{-2}}{(2\pi )^{(p+3)/2}V_{q-1}}\sum_{n=1}^{\infty }%
\frac{\cos (2\pi n\alpha _{p+1})}{(nL_{p+1})^{p+3-2b}}  \notag \\
&&\times \sum_{\mathbf{n}_{q-1}\in \mathbf{Z}^{q-1}}f_{(p+3)/2-b}(nL_{p+1}k_{%
\mathbf{n}_{q-1}}),\;i=1,2,\ldots p,  \label{T00minAs}
\end{eqnarray}%
and for the stresses along the compact dimensions%
\begin{eqnarray}
\Delta _{p+1}\langle T_{p+1}^{p+1}\rangle _{p,q} &\approx &\frac{p+1-b}{b-1}%
\Delta _{p+1}\langle T_{0}^{0}\rangle _{p,q}+\frac{2^{b+1}\Gamma (b)A\Omega
^{-2}}{(2\pi )^{(p+3)/2}V_{q-1}}\sum_{n=1}^{\infty }\frac{\cos (2\pi n\alpha
_{p+1})}{(nL_{p+1})^{p+2-2b}}  \notag \\
&&\times \sum_{\mathbf{n}_{q-1}\in \mathbf{Z}^{q-1}}(nL_{p+1}k_{\mathbf{n}%
_{q}-1})^{2}f_{(p+1)/2-b}(nL_{p+1}k_{\mathbf{n}_{q-1}}),  \notag \\
\Delta _{p+1}\langle T_{i}^{i}\rangle _{p,q} &\approx &\Delta _{p+1}\langle
T_{0}^{0}\rangle _{p,q}-\frac{2^{b+1}\Gamma (b)A\Omega ^{-2}}{(2\pi
)^{(p+3)/2}V_{q-1}}\sum_{n=1}^{\infty }\frac{\cos (2\pi n\alpha _{p+1})}{%
(nL_{p+1})^{p+2-2b}}  \notag \\
&&\,\times \sum_{\mathbf{n}_{q-1}\in \mathbf{Z}%
^{q-1}}(nL_{p+1}k_{i})^{2}f_{(p+1)/2-b}(nL_{p+1}k_{\mathbf{n}%
_{q-1}}),\;ip+2,\ldots ,D.  \label{TiiminAs}
\end{eqnarray}%
Hence, in this case the topological part behaves as $\Delta _{p+1}\langle
T_{i}^{i}\rangle _{p,q}\propto t^{-2c}$ in the limit $t\rightarrow \infty $
for $c>1$ and in the limit $t\rightarrow 0$ for $0<c<1/D$.

For small values of the ratio $\eta /L_{p+1}$ and imaginary $\nu $ we use
the asymptotic formula (\ref{KIapp}). To the leading order this gives (no
summation over $l$)%
\begin{equation}
\Delta _{p+1}\langle T_{l}^{l}\rangle _{p,q}\approx \frac{%
2^{(1-p)/2}B_{l}\Omega ^{-D-1}}{\pi ^{(p+3)/2}V_{q}L_{p+1}^{p}\eta }\cos %
\left[ 2|\nu |\ln (L_{p+1}/\eta )+\phi _{l}\right] ,\;L_{p+1}/\eta \gg 1,
\label{Tlllargeim}
\end{equation}%
where the constants $B_{l}$ and $\phi _{l}$ are defined by the relation
\begin{equation}
B_{l}e^{i\phi _{l}}=2^{i|\nu |}\Gamma (i|\nu |)F^{(l)}\sum_{\mathbf{n}%
_{q-1}\in \mathbf{Z}^{q-1}}\sum_{n=1}^{\infty }\frac{\cos (2\pi n\alpha
_{p+1})}{n^{p+1-2i|\nu |}}f_{(p+1)/2-i|\nu |}(nL_{p+1}k_{\mathbf{n}_{q-1}}),
\label{Bl}
\end{equation}%
with $F^{(l)}$, $l=0,1,\ldots ,D$, from (\ref{F0l}). Hence, for imaginary $%
\nu $ the behaviour of the VEVs is oscillatory. The corresponding amplitude
behaves as $t^{-cD-1}$ and we have damping oscillations.

Summarizing the analysis given above, for a minimally coupled scalar field
in the models with $0<c<1$, at late stages, $t\rightarrow \infty $, the
topological part in the VEV of the energy-momentum tensor behaves as $%
t^{-(D+1)c}$. At early stages, $t\rightarrow 0$, we have $\langle
T_{i}^{i}\rangle _{p,q}^{\mathrm{(t)}}\propto t^{-2c}$ for $0<c<1/D$ and $%
\langle T_{i}^{i}\rangle _{p,q}^{\mathrm{(t)}}\propto t^{-2Dc}$ for $1/D<c<1$%
. In the models with $c>1$ and at late stages of the cosmological expansion
the topological part decays like $t^{-2c}$. At early stages and for $c>1$
one has $\langle T_{i}^{i}\rangle _{p,q}^{\mathrm{(t)}}\propto t^{-(D+1)c}$.

\section{Special case and numerical examples}

\label{sec:Special}

As an application of the general formulae given above let us consider the
special case of topology $R^{D-1}\times S^{1}$. For this case the expression
for the topological part in the VEV of the field squared takes the form
\begin{eqnarray}
\langle \varphi ^{2}\rangle _{D-1,1}^{\mathrm{(t)}} &=&\frac{4(2\pi
)^{-D/2-1}}{(\Omega \eta )^{D-1}}\sum_{n=1}^{\infty }\frac{\cos (2\pi
n\alpha _{D})}{(nL_{D}/\eta )^{D-2}}  \notag \\
&&\times \int_{0}^{\infty }dyy\left[ I_{-\nu }(y)+I_{\nu }(y)\right] K_{\nu
}(y)f_{D/2-1}(nyL_{D}/\eta ).  \label{phi2Sp}
\end{eqnarray}%
Note that in this formula $\Omega \eta =t/|1-c|$. For a conformally coupled
field this formula is reduced to%
\begin{equation}
\langle \varphi ^{2}\rangle _{D-1,1}^{\mathrm{(t)}}=\frac{\Gamma ((D-1)/2)}{%
2\pi ^{(D+1)/2}(\Omega L_{D})^{D-1}}\sum_{n=1}^{\infty }\frac{\cos (2\pi
n\alpha _{D})}{n^{D-1}},\;\xi =\xi _{D}.  \label{phi2Spconf}
\end{equation}%
Formula (\ref{phi2Spconf}) describes the asymptotic behavior of the VEV in
the general case of $\xi $ in early stages of the cosmological expansion ($%
t\rightarrow 0$) for $c>1$ and at late stages ($t\rightarrow \infty $) for $%
0<c<1$. In the opposite limit of late times ($t\rightarrow \infty $) for $c>1
$ and early times ($t\rightarrow 0$) for $0<c<1$, for positive values of $%
\nu $, the asymptotic behavior is given by the expression%
\begin{equation}
\langle \varphi ^{2}\rangle _{D-1,1}^{\mathrm{(t)}}\approx \Gamma (D/2-\nu )%
\frac{\Gamma (\nu )(L_{D}/\eta )^{2\nu -1}}{2\pi ^{D/2+1}(\Omega L_{D})^{D-1}%
}\sum_{n=1}^{\infty }\frac{\cos (2\pi n\alpha _{D})}{n^{D-2\nu }}.
\label{phi2SpAs}
\end{equation}%
Note that in this expression $\Omega L_{D}$ is the comoving length of the
compact dimension. For imaginary values of $\nu $ we have the asymptotic
described by the relation (\ref{Tlllargeim}). In particular, in the models
of power law inflation ($c>1$) we have $\langle \varphi ^{2}\rangle
_{D-1,1}^{\mathrm{(t)}}\propto t^{(1-D)c}$ at early times. At late times the
topological part behaves monotonically as $t^{2(b-\nu )(1-c)}$ in the case
of positive $\nu $ and oscillatory like $t^{1-cD}\cos [2|\nu |(c-1)\ln t+%
\mathrm{const}]$ for imaginary values of $\nu $. In the latter case the
distance between the nearest zeros linearly increases with increasing $t$.

\begin{figure}[tbph]
\begin{center}
\epsfig{figure=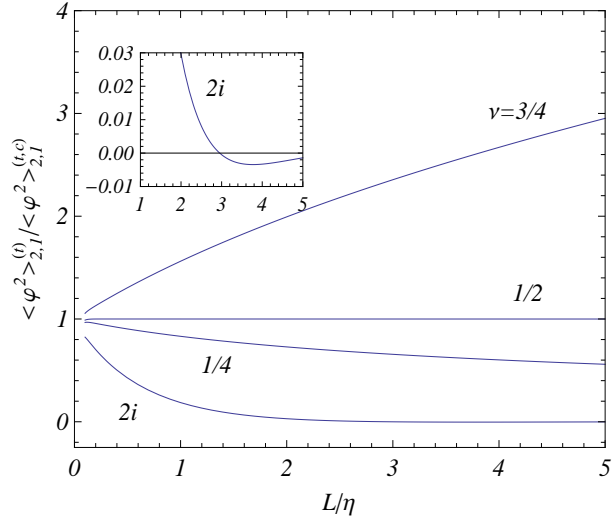,width=8cm,height=7.cm}
\end{center}
\caption{The ratio $\langle \protect\varphi ^{2}\rangle _{D-1,1}^{\mathrm{(t)%
}}/\langle \protect\varphi ^{2}\rangle _{D-1,1}^{\mathrm{(t,c)}}$ versus $L/%
\protect\eta $ for different values of the parameter $\protect\nu $ (numbers
near the curves) and for $D=3$ scalar field with the periodicity condition
along the compact dimension.}
\label{fig1}
\end{figure}

Note that the ratio $\langle \varphi ^{2}\rangle _{D-1,1}^{\mathrm{(t)}%
}/\langle \varphi ^{2}\rangle _{D-1,1}^{\mathrm{(t,c)}}$, with $\langle
\varphi ^{2}\rangle _{D-1,1}^{\mathrm{(t,c)}}$ being the VEV for a
conformally coupled field given by (\ref{phi2Spconf}), is a function of $%
L_{D}/\eta $ alone and depends on the parameters $\xi $ and $c$ through $\nu
$. In figure \ref{fig1}, we have plotted this ratio as a function of $L/\eta
$, with $L=L_{D}$ being the length of the compact dimension, for untwisted $%
D=3$ scalar field ($\alpha _{D}=0$) and for various values of the parameter $%
\nu $ (numbers near the curves). In this special case one has $\langle
\varphi ^{2}\rangle _{D-1,1}^{\mathrm{(t,c)}}=1/[12(\Omega L_{D})^{2}]$.
Note that the ratio $L/\eta $ is related to the comoving length of the
compact dimension, measured in units of the Hubble length, by Eq. (\ref{LcrH}%
). Figure \ref{fig1} clearly shows that the adiabatic approximation for the
topological part is valid only for small values of the ratio $L^{\mathrm{(c)}%
}/r_{\mathrm{H}}$.

In the special case under consideration, for the topological part in the VEV
of the energy-momentum tensor we have the following representation (no
summation over $i$)
\begin{eqnarray}
\langle T_{i}^{i}\rangle _{D-1,1}^{\mathrm{(t)}} &=&\frac{4(2\pi )^{-D/2-1}}{%
(\Omega \eta )^{D+1}}\sum_{n=1}^{\infty }\frac{\cos (2\pi n\alpha _{D})}{%
(nL_{D}/\eta )^{D-2}}\int_{0}^{\infty }dy\,y^{3-2b}  \notag \\
&&\times \left[ F^{(i)}(y)f_{D/2-1}(nyL/\eta )-\,\frac{\widetilde{I}_{\nu
}(y)\widetilde{K}_{\nu }(y)}{(nyL_{D}/\eta )^{2}}f_{D-1}^{(i)}(nyL_{D}/\eta )%
\right] ,  \label{TiiSp}
\end{eqnarray}%
where $F^{(0)}(y)$ is defined by (\ref{F0}), $f_{D-1}^{(0)}(x)=0$, and $%
F^{(i)}(y)=F(y)$ for $i=1,\ldots ,D$. The functions $f_{D-1}^{(i)}(x)$ for $%
i=1,\ldots ,D$ are given by expressions (\ref{fp}). For a conformally
coupled field we have (no summation over $i=0,1,\ldots ,D-1$)%
\begin{equation}
\langle T_{i}^{i}\rangle _{D-1,1}^{\mathrm{(t)}}=-\frac{1}{D}\langle
T_{D}^{D}\rangle _{D-1,1}^{\mathrm{(t)}}=-\frac{\Gamma ((D+1)/2)}{\pi
^{(D+1)/2}(\Omega L_{D})^{D+1}}\sum_{n=1}^{\infty }\frac{\cos (2\pi n\alpha
_{D})}{n^{D+1}}.  \label{TiiSpconf}
\end{equation}%
Formula (\ref{TiiSpconf}) describes the asymptotic behavior of the VEV for
the general case of the curvature coupling parameter in the limit $%
L_{D}/\eta \ll 1$. In the opposite limit the corresponding asymptotic
formulae directly follow from the expressions (\ref{T00largeL}) and (\ref%
{Tlllargeim}) for real and imaginary $\nu $ respectively. The formulae are
simplified in the case $c=2/(D+1)$ for general curvature coupling parameter
(no summation over $i$):%
\begin{equation}
\langle T_{i}^{i}\rangle _{D-1,1}^{\mathrm{(t)}}=\frac{\Gamma ((D-1)/2)}{\pi
^{(D+1)/2}(\Omega L_{D})^{D+1}}\sum_{n=1}^{\infty }\frac{\cos (2\pi n\alpha
_{D})}{n^{D+1}}\left[ f_{i}\frac{\xi -\xi _{D}}{D-1}\left( \frac{nL_{D}}{%
\eta }\right) ^{2}-g_{i}\frac{D-1}{2}\right] ,  \label{TiiSpSp}
\end{equation}%
where $g_{i}=1$ for $i=0,1,\ldots ,D-1$, $g_{D}=-D$ and $f_{i}$ is defined
after formula (\ref{TiiSpc}).

\begin{figure}[tbph]
\begin{center}
\begin{tabular}{cc}
\epsfig{figure=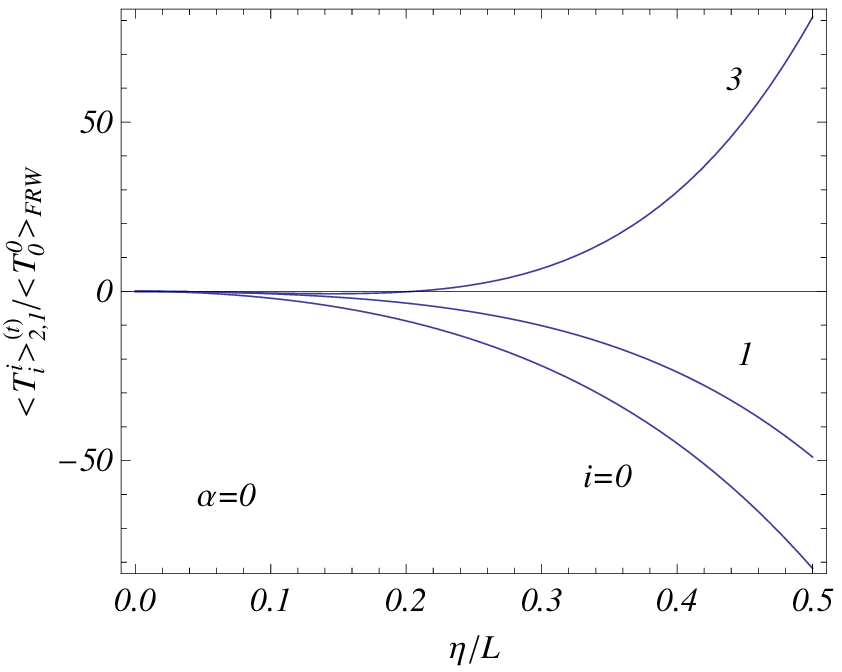,width=7cm,height=6.cm} & \quad %
\epsfig{figure=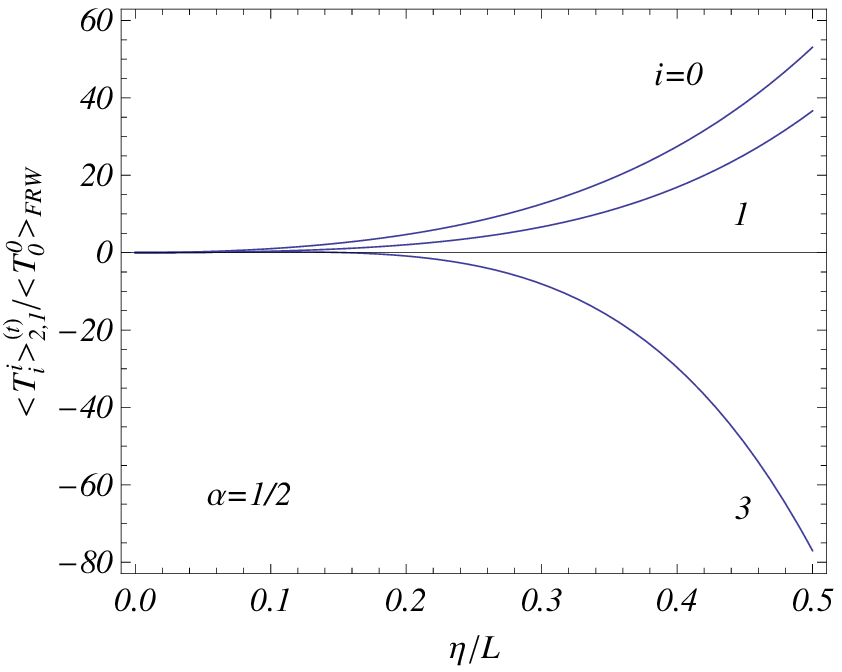,width=7cm,height=6cm}%
\end{tabular}%
\end{center}
\caption{The topological parts in the VEV of the energy-momentum
tensor for the case of radiation driven model with spatial
topology $R^{2}\times S^{1}$ as functions of the ratio
$\protect\eta /L$. The left/right panel corresponds to
untwisted/twisted scalar fields. } \label{fig2}
\end{figure}

In order to see the relative contributions of separate terms in the VEV of
the energy-momentum tensor, in figure \ref{fig2} we have plotted the ratio $%
\langle T_{i}^{i}\rangle _{D-1,1}^{\mathrm{(t)}}/\langle T_{0}^{0}\rangle _{%
\mathrm{FRW}}$ in the radiation driven model ($c=2/(D+1)$) for $D=3$
minimally coupled untwisted (left panel) and twisted (right panel) scalar
fields versus $\eta /L$ with $L=L_{D}$ being the length of the compact
dimension. Here, $\langle T_{0}^{0}\rangle _{\mathrm{FRW}}=1/[480\pi
^{2}(2t)^{4}]$ is the vacuum energy density for a massless minimally coupled
scalar field in the corresponding FRW model with trivial topology. Note that
for the special case under consideration one has $\eta =(2/\alpha )t^{1/2}$.
In the same model and for a conformally coupled scalar fields we have (no
summation over $i=1,2$)%
\begin{equation}
\langle T_{0}^{0}\rangle _{2,1}^{\mathrm{(t)}}=\langle T_{i}^{i}\rangle
_{2,1}^{\mathrm{(t)}}=-\frac{1}{3}\langle T_{3}^{3}\rangle _{2,1}^{\mathrm{%
(t)}}=\frac{\pi ^{2}C}{(2t)^{4}}\left( \frac{\eta }{L}\right) ^{4},
\label{TD3conf}
\end{equation}%
where $C=-1/90$ and $C=7/720$ for untwisted and twisted fields respectively.

Now let us estimate the relative contributions of different sources in the
total energy density for the $D=3$ radiation dominated model. Introducing
the Planck mass $M_{\mathrm{P}}^{2}=1/G$ and the Planck time $t_{\mathrm{P}}$%
, for the energy density being the source of the background metric one has $%
\rho \sim $ $M_{\mathrm{P}}^{4}/(t/t_{\mathrm{P}})^{2}$. The average vacuum
energy density in FRW model with trivial topology is estimated as $\langle
T_{0}^{0}\rangle _{\mathrm{FRW}}\sim M_{\mathrm{P}}^{4}/(t/t_{\mathrm{P}%
})^{4}$. For the topological part in the vacuum energy density in the case
of the conformally coupled field we have $\langle T_{0}^{0}\rangle _{\mathrm{%
conf}}^{\mathrm{(t)}}\sim M_{\mathrm{P}}^{4}/(L^{\mathrm{(c)}}/l_{\mathrm{P}%
})^{4}$, where $L^{\mathrm{(c)}}=a(t)L$ is the comoving length of the
compact dimension and $l_{\mathrm{P}}$ is the Planck length. For a minimally
coupled field one has $\langle T_{0}^{0}\rangle _{\mathrm{\min }}^{\mathrm{%
(t)}}\sim \langle T_{0}^{0}\rangle _{\mathrm{conf}}^{\mathrm{(t)}}[1+\mathrm{%
const}\cdot (L^{\mathrm{(c)}}/t)^{2}]$. From these relations, for the ratios
of the separate parts in the vacuum energy density to the source term we find%
\begin{equation}
\langle T_{0}^{0}\rangle _{\mathrm{FRW}}/\rho \sim \left( t_{\mathrm{P}%
}/t\right) ^{2},\;\langle T_{0}^{0}\rangle _{\mathrm{conf}}^{\mathrm{(t)}%
}/\rho \sim \left( l_{\mathrm{P}}r_{\mathrm{H}}/L^{\mathrm{(c)}2}\right)
^{2},\;  \label{relcont}
\end{equation}%
where $r_{\mathrm{H}}$ is the Hubble length. From these relations we
conclude that the term $\langle T_{0}^{0}\rangle _{\mathrm{FRW}}$ is
comparable to the source term $\rho $ only near the Planck time. The
topological part dominates the part $\langle T_{0}^{0}\rangle _{\mathrm{FRW}%
} $ when $L^{\mathrm{(c)}}<r_{\mathrm{H}}$ and becomes of the order of the
source term for $L^{\mathrm{(c)}2}\sim l_{\mathrm{P}}r_{\mathrm{H}}$. In the
latter case the backreaction of quantum topological effects should be taken
into account.

\section{Conclusion}

\label{sec:Conc}

In the present paper we have investigated one-loop quantum effects for a
scalar field with general curvature coupling, induced by toroidal
compactification of spatial dimensions in spatially flat FRW cosmological
models with power law scale factor. We treat gravity as a given classical
background field and do not consider the backreaction of the quantum effects
on the metric. General boundary conditions with arbitrary phases are
considered along compact dimensions. As special cases they include
periodicity and antiperiodicity conditions corresponding to untwisted and
twisted fields. The boundary conditions imposed on possible field
configurations change the spectrum of vacuum fluctuations and lead to the
Casimir-type contributions in the VEVs of physical observables. Among the
most important characteristics of the vacuum state are the expectation
values of the field squared and the energy-momentum tensor. Though the
corresponding operators are local, due to the global nature of the vacuum
these VEVs carry an important information on the global structure of the
background spacetime. As the first step we evaluate the positive-frequency
Wightman function. By using the Abel-Plana summation formula, we have
derived recurrence relation connecting the Wightman functions for the
topologies $R^{p}\times (S^{1})^{q}$ and $R^{p+1}\times (S^{1})^{q-1}$. The
repeating application of this relation allows to present the Wightman
function as the sum of the function for topologically trivial FRW model and
the topological part. The latter is finite in the coincidence limit and in
this way the renormalization of the VEVs for the field squared and the
energy-momentum tensor is reduced to that for the FRW universe with trivial
topology. The latter problem is well investigated in the literature and we
concentrate on the topological parts.

The topological parts are given by formulae (\ref{phi2}) and (\ref%
{phi2decomp2}) for the field squared and by formulae (\ref{T00}), (\ref{Tiin}%
), (\ref{TikDecomp}) for the energy density and the stresses. In the case of
a conformally coupled scalar field the problem is conformally related to the
corresponding problem in flat spacetime with topology $R^{p}\times
(S^{1})^{q}$ and the general formulae are simplified to (\ref{phi21}) and (%
\ref{TiiConf}) for the field squared and the energy-momentum tensor
respectively. These formulae describe the asymptotic behavior of the VEVs in
the general case of the curvature coupling in the limit $L_{l}/\eta \ll 1$.
In this limit the comoving lengths of the compact dimensions measured in
units of the Hubble length are small and it corresponds to early stages of
the cosmological expansion for $0<c<1$ and to late stages of the
cosmological expansion when $c>1$. The topological parts in the VEVs behave
like $\langle \varphi ^{2}\rangle _{p,q}^{\mathrm{(t)}}\propto t^{c(1-D)}$
and $\langle T_{l}^{k}\rangle _{p,q}^{\mathrm{(t)}}\propto t^{-c(1+D)}$ and
the stresses along the uncompactified dimensions are equal to the vacuum
energy density.

In the opposite limit $L_{l}/\eta \gg 1$, corresponding to large
comoving lengths of compact dimensions compared to the Hubble
length, the behaviour of the VEVs is qualitatively different for
real and imaginary values of the parameter $\nu $ defined by Eq.
(\ref{nu}). For real values, the asymptotics are given by formulae
(\ref{phi2large}) and (\ref{T00largeL}) and the
topological parts in the VEVs behave as $\langle \varphi ^{2}\rangle _{p,q}^{%
\mathrm{(t)}}\propto t^{2(c-1)\nu -cD+1}$ and $\langle T_{l}^{k}\rangle
_{p,q}^{\mathrm{(t)}}\propto t^{2(c-1)\nu -cD-1}$. In this limit the vacuum
stresses are isotropic and the equation of state for the topological parts
in the vacuum energy density and pressures is of the barotropic type. For a
minimally coupled scalar field in the case $c\in (0,1/D)\cup (1,\infty )$
and for a conformally coupled field, the leading terms in the asymptotic
expansions for the VEV\ of the energy-momentum tensor vanish. The
corresponding asymptotics are given by $\langle T_{l}^{k}\rangle _{p,q}^{%
\mathrm{(t)}}\propto t^{-c(1+D)}$ for conformally coupled field and by $%
\langle T_{l}^{k}\rangle _{p,q}^{\mathrm{(t)}}\propto t^{-2c}$ for a
minimally coupled field in the model with $c\in (0,1/D)\cup (1,\infty )$
(see formulae (\ref{T00minAs}) and (\ref{TiiminAs})). In these special cases
the vacuum stresses are anisotropic though the equation of state remains of
the barotropic type with the coefficients depending on the lengths of
compact dimensions. In the limit $L_{l}/\eta \gg 1$ and for imaginary values
of the parameter $\nu $ the asymptotic behaviour of the topological parts is
described by the formulae (\ref{phi2largeim}) and (\ref{Tlllargeim}) and
this behaviour is oscillatory with the oscillating factor in the form $\cos
[2|\nu |(c-1)\ln (t/t_{0})+\psi ]$. For the field squared the amplitude of
the oscillations behaves as $t^{1-cD}$. For the energy-momentum tensor the
amplitude of the oscillations decreases like $t^{-cD-1}$ and the
oscillations are damping for all values of the parameter $c$.

The general results for the VEVs of the field squared and the
energy-momentum tensor are specified in section \ref{sec:Special} for the
case of a model with a single compact dimension. By numerical examples for
this case we have illustrated that the adiabatic approximation for
nonconformally coupled fields is valid only in the limit when the comoving
length of the compact dimension is very short compared to the Hubble length.
We have also estimated the relative contributions of various sources in the
total energy density. In particular, in the radiation driven model, for the
comoving lengths $L^{\mathrm{(c)}}\lesssim l_{\mathrm{P}}r_{\mathrm{H}}$ the
topological part of the energy density becomes of the order of the source
term and the backreaction of quantum effects should be taken into account.

In this paper we have considered the case of a massless field when the
equation (\ref{Eqphi}) for the mode functions is exactly integrable in the
class of special functions. By using this equation, we can estimate the role
of effects due to the nonzero mass of the field. The ratio of the first and
second terms in figure braces of (\ref{Eqphi}) is of the order $m^{2}t^{2}$.
Hence, for $mt\ll 1$ the effects related to the nonzero mass may be
disregarded and the results described in the present paper are applicable
for massive fields as well.

\section*{Acknowledgments}

The work was supported by the Armenian Ministry of Education and Science
Grant No. 0124. A.A.S. gratefully acknowledges the hospitality of the Abdus
Salam International Centre for Theoretical Physics (Trieste, Italy) where
part of this work was done.

\end{document}